\documentstyle[prl,aps,epsf,twocolumn]{revtex}
\begin{document}
\newcommand{\bo}{\={o} }
\newcommand{\be}{\begin{equation} }
\newcommand{\ee}{\end{equation} }

\title{Understanding conserved amino acids in proteins}
\author{Nikolay V. Dokholyan, Leonid A. Mirny and Eugene I. Shakhnovich}
\address{Department of Chemistry, Harvard University, 12 Oxford
Street, Cambridge, MA 02138, USA}

\date{\today}

\maketitle

{\bf It has been conjectured that evolution exerted pressure to
preserve amino acids bearing thermodynamic, kinetic, and functional
roles. In this letter we show that the physical requirement to
maintain protein stability gives rise to a sequence conservatism
pattern that is in remarkable agreement with that found in natural
proteins. Based on the physical properties of amino acids, we propose
a model of evolution that explains conserved amino acids across
protein families sharing the same fold.}

Molecular evolution is a sumptuous natural laboratory that provides an
invaluable source of information about the structure, dynamics and
function(s) of biomolecules. This information has already been widely
used to understand the folding kinetics, thermodynamics, and function
of proteins (e.g. \cite{Branden98,Mirny99}). The basic belief behind
the majority of such studies is that evolution optimizes, to a certain
extent, the properties of proteins, so that they become more
sufficiently stable, and have better folding and functional
properties.

Recent studies \cite{Mirny99,Ptitsyn98} identified positions in
several common protein folds where amino acids are universally
conserved within each family of proteins having that fold. Such
positions are localized in structure, and their unusually strong
conservatism may be due to functional reason (e.g. super-site), or
folding kinetics (folding nucleus)
\cite{Mirny98,vanNuland98a,vanNuland98b,Dokholyan00a}. In contrast to
function and folding kinetics, evolutionary pressure to maintain
stability may be applied ``more evenly'' because all amino acids
contribute, to a lesser or greater extent, to protein stability via
their interaction with other amino acid residues and with water.

In this letter we develop a model that provides a rationale for
conservatism patterns caused by selection for stability. Our model is
of {\em equilibrium} evolution that maintains stability and other
properties achieved at an earlier, prebiotic stage. To this end we
propose that stability selection accepts only those mutations that
keep energy of the native protein, $E$, below a certain threshold
$E_0$ necessary to maintain an energy gap
\cite{Sali94,Shakhnovich93,Ramanathan94,Finkelstein95}.  The
requirement to maintain an energy threshold for the viable sequences
makes the equilibrium ensemble of sequences analogous to a
microcanonical ensemble. In analogy with statistical mechanics, a more
convenient and realistic description of the sequence ensemble is a
canonical ensemble, whereby strict requirements on energy of the
native state is replaced by a ``soft'' evolutionary pressure that
allows energy fluctuations from sequence to sequence but makes
sequences with high energy in the native state unlikely.  In the
canonical ensemble of sequences, the probability of finding a
particular sequence, $\{\sigma\}$, in the ensemble follows the
Boltzmann distribution
\cite{Shakhnovich93,Ramanathan94,Finkelstein95,Pande95}
\be
P(\{\sigma\})=\frac{\exp(-E\{\sigma\}/T)}{Z}\, , 
\ee
where $T$ is the effective temperature of the canonical ensemble of
sequences that serves as a measure of evolutionary pressure and $Z =
\sum_{\{\sigma\}} \exp{( -E\{\sigma\}/T)} $ is the partition function
taken in sequence space.

Next, we apply a mean-field approximation that replaces all
multiparticle interactions between amino acids by interaction of each
amino acid with an effective field $\Phi$ acting on this amino acid
from the rest of the protein. This approximation presents
$P(\{\sigma\})$ in a multiplicative form as $\prod_{k=1}^{N}
p(\sigma_k)$ of probabilities to find an amino acid $\sigma$ at
position $k$. $p(\sigma_k)$ also obeys Boltzmann statistics
\be
p(\sigma_k) = \frac{\exp(-\Phi(\sigma_k)/T)}{\sum_{\sigma}
\exp(-\Phi(\sigma_k)/T)}\, .
\label{eq:pmf}
\ee
The mean field potential $\Phi(\sigma_k)$ is the effective
potential energy between amino acid $\sigma$ and all amino acids
interacting with it, i.~e.
\be
\Phi(\sigma_k)=\sum_{i\neq k }^{N} U(\sigma_k, \sigma_i)\Delta_{ki}\, .
\label{eq:phi}
\ee
Matrix $U$ describes energy parameters in contact approximation and
matrix $\Delta$ is a contact matrix for protein native structure (see
{\sf Methods} for more detail). The potential $\Phi$ is similar in
spirit to the protein profile introduced by Bowie et
al. \cite{Bowie91} to identify protein sequences that fold into a
specific 3D structure.

For each member, $m$, of the fold family (FSSP database \cite{Holm93})
we compute the mean-field probability, $p_m(\sigma_k)$, using
Eq.~(\ref{eq:pmf}) and then, we compute the average mean-field
probability,
\be
p_{MF}(\sigma_k) = \frac{1}{N_s} \sum_{m=1}^{N_s} p_m(\sigma_k)\, .  
\label{eq:pmf1}
\ee
Eqs. (\ref{eq:pmf}) --- (\ref{eq:pmf1}), along with properly selected
energy function, $U$, make it possible to predict probabilities of all
amino acid types and sequence entropy $S_{MF}(k)$ at each position $k$
\be
S_{MF}(k)=- \sum_{\sigma}p_{MF}(\sigma_k)\ln p_{MF}(\sigma_k)\, 
\label{eq:SMF}
\ee
from the native structure of a protein. The summation is taken over
all possible values of $\sigma$.

Theoretical predictions from statistical-mechanical analysis can be
compared with data on real proteins. In order to determine
conservatism in real proteins we note that the space of sequences that
fold into the same protein structure presents a two-tier system, where
homologous sequences are grouped into families and there is no
recognizable sequence homology between families despite the fact that
they fold into closely related structures
\cite{Mirny99,Rost97,Tiana00}.

Using the database of protein families with close sequence similarity
(HSSP database \cite{Dodge98}), we compute frequencies of amino acids
at each position, $k$, of aligned sequences, $P_m(\sigma_k)$, for a
given, $m$th, family of proteins. We average these frequencies {\it
across} all $N_s$ families sharing the same fold that are present in
FSSP database \cite{Holm93}:
\be
P_{acr}(\sigma_k) = \frac{1}{N_s}\sum_{m=1}^{N_s} P_m(\sigma_k)\, .
\ee
Next, we determine the sequence entropy, $S_{acr}(k)$, at each
position, $k$, of structurally aligned protein analogs:
\be
S_{acr}(k) = - \sum_{\sigma} P_{acr}(\sigma_k)\ln P_{acr}(\sigma_k) \, .
\label{eq:s1}
\ee

If the stability selection were a factor in evolution of proteins and
our model captures it, then we should observe correlation between
predicted mean field based sequence entropies, $S_{MF}(k)$, and actual
sequence entropies $S_{acr}(k)$ in real proteins.  Thus, the question
is: ``Can we find such $T$, so that predicted conservatism profile
$S_{MF}(k)$ matches the real one $S_{acr}(k)$?''

By varying the values of the temperature, $T$ in the range $0.1\leq
T\leq 4.0$, we minimize the distance, $D^2\equiv\sum_{k=1}^N
(S_{MF}(k) - S_{acr}(k))^2$, between the predicted and observed
conservatism profiles. We exclude from this sum such positions in
structurally aligned sequences that have more than 50\% gaps in the
structural (FSSP) alignment. We denote by $T_{sel}$ the temperature
that minimizes $D$.

\begin{figure}[hbt]
\centerline{ \vbox{ \hbox{\epsfxsize=10.0cm
\epsfbox{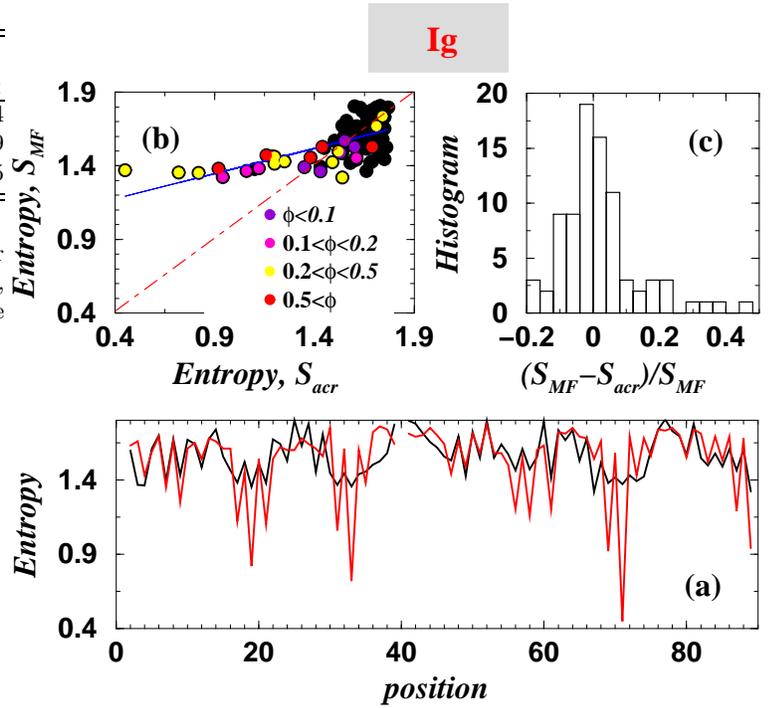}  }}}
\vspace*{0.5cm}
\caption{(a) The values $S_{MF}(k)$ (black line) and $S_{acr}(k)$ (red
line) for all positions, $k$, for the Ig-fold. The lower the values of
$S_{MF}(k)$ the more conservative amino acids are at these positions.
(b) The scatter plot of predicted $S_{MF}(k)$ versus observed
$S_{acr}(k)$. The linear regression correlation coefficients are shown
in Table~\protect\ref{t:1}. Blue line is the linear regression has the
slope different than 1 (red line), corresponding to the $S_{MF}(k) =
S_{acr}(k)$ relation. (c) The histogram of the differences between
$S_{MF}(k)$ and $S_{acr}(k)$.  In (b) we assign colors to data points
corresponding to amino acids with the specific range of $\phi$-values
\protect\cite{Hamill00}: red, if $0.5<\phi<1$, yellow, if
$0.2<\phi<0.5$, magenta, if $0.1<\phi<0.2$, violet if $\phi<0.1$, and
black if $\phi$-values are not determined.}
\label{fig:1}
\end{figure}

\begin{figure}[hbt]
\centerline{ \vbox{ \hbox{\epsfxsize=10.0cm
\epsfbox{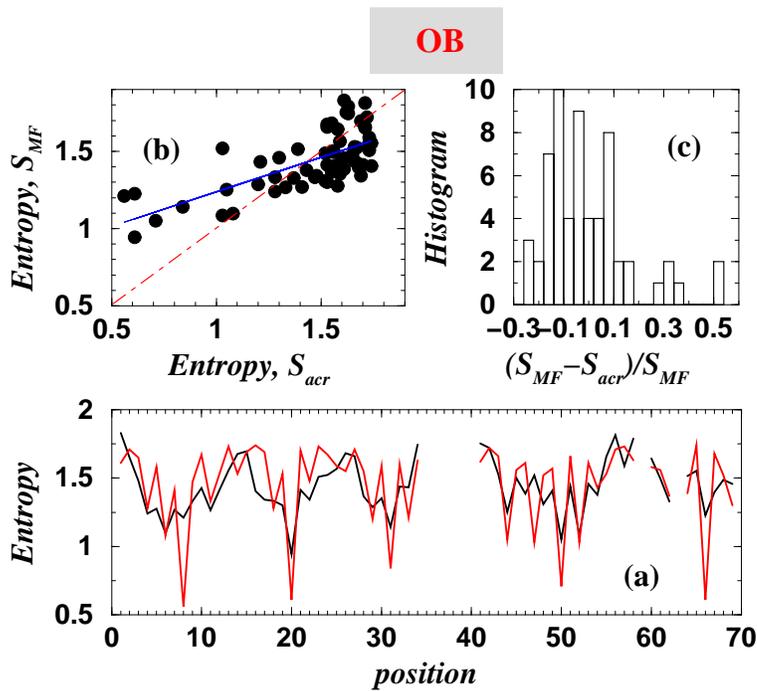}  }}}
\vspace*{0.5cm}
\caption{(a) --- (c) The same as Fig.\protect\ref{fig:1} but for the
OB-fold.}
\label{fig:2}
\end{figure}

\begin{figure}[hbt]
\centerline{ \vbox{ \hbox{\epsfxsize=10.0cm
\epsfbox{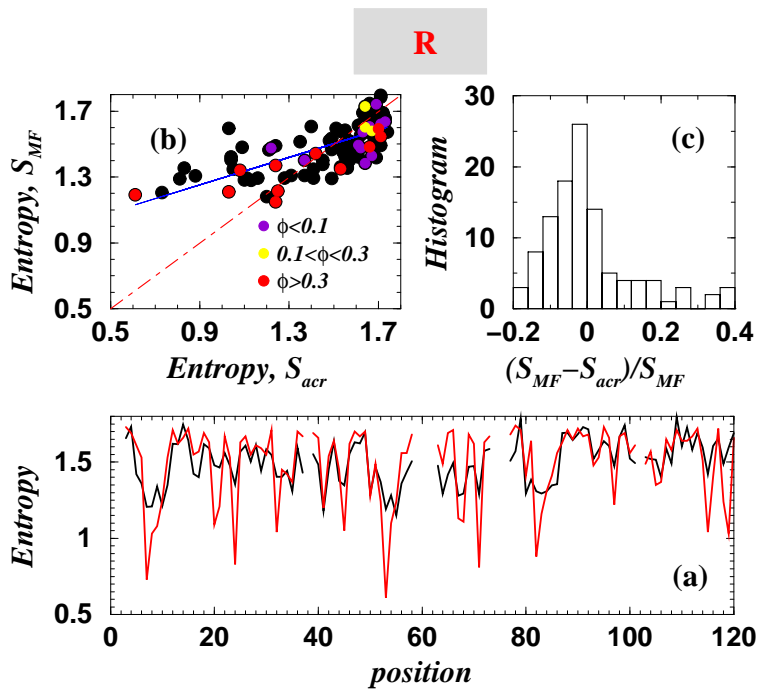}  }}}
\vspace*{0.5cm}
\caption{(a) --- (c) The same as Fig.\protect\ref{fig:1} but for
R-fold.  In (b) we assign colors to data points corresponding to amino
acids with the specific range of $\phi$-values \protect\cite{Lopez96}:
red, if $0.3<\phi<1$, yellow, if $0.1<\phi<0.3$, violet if $\phi<0.1$,
and black if $\phi$-values are not determined.}
\label{fig:3}
\end{figure}

We study three folds: Immunoglobulin fold (Ig),
Oligonucleotide-binding fold (OB), and Rossman fold (R). We compute
correlation coefficient \cite{Press89} between values of $S_{MF}(k)$,
obtained at $T_{sel}$, and $S_{acr}(k)$ for all three folds. The
results are summarized in Table~\ref{t:1}.  The plots of $S_{MF}(k)$
and $S_{acr}(k)$ versus $k$ as well as their scatter plots are shown
in Figs.~\ref{fig:1}--\ref{fig:3}(a,b).

The correlation between $S_{MF}(k)$ and $S_{acr}(k)$ is remarkable for
all three folds and indicates that our mean field model is able to
select the conserved amino acids in protein fold families.  It is
fully expected that the correlation coefficient is smaller than 1. The
reason for this is that computation of $S_{MF}(k)$ takes into account
evolutionary selection for stability only and it does not take into
account possible additional pressure to optimize kinetic or functional
properties.

The additional evolutionary pressure due to kinetic or functional
importance of amino acids results in pronounced deviations of $S_{MF}$
from $S_{acr}$ for few amino acids that may be kinetically or
functionally important. A number of amino acids whose conservatism is
much greater than predicted by our model form a group of ``outliers'',
from otherwise very close correspondence between $S_{MF}$ and
$S_{acr}$. To demonstrate that some of those amino acids are important
for folding kinetics and as such they can be under additional
evolutionary pressure, we color data points on $S_{MF}$ versus
$S_{acr}$ scatter plot according to the range of $\phi$-values
\cite{Itzhaki95} that corresponding amino acids fall
into. Thermodynamics and kinetic role of individual amino acids was
studied extensively {\it (i)} by Hamill et al.  \cite{Hamill00} for
the TNfn3 (1TEN) protein, and {\it (ii)} by L\'opez-Hern\'andez and
Serrano \cite{Lopez96} for the CheY protein. We use the $\phi$-values
for individual amino acids obtained in \cite{Hamill00,Lopez96}. We
observe that {\it (i)} for TNfn3 protein most of the points on
Fig.~\ref{fig:1}(b) that belong to the outlier group have
$\phi$-values ranging from 0.2 to 1, and {\it (ii)} for CheY protein
most of the points (for which $\phi$-values are known) on
Fig.~\ref{fig:3}(b) that belong to the outlier have $\phi$-values
ranging from 0.3 to 1.

To conclude, we presented a theory that explains sequence conservation
caused by the most basic and universal evolutionary pressure in
proteins to maintain stability. The theory predicts very well sequence
entropy for the majority of amino acids, but not all of them. The
amino acids that exhibit considerably higher conservatism than
predicted from stability pressure alone are likely to be important for
function and/or folding. Comparison of the ``base-level'' stability
conservatism $S_{MF}(k)$ with $S_{acr}(k)$ - actual conservatism
profile of a protein fold - allows to identify functionally and
kinetically important amino acid residues and potentially gain
specific insights into folding and function of a protein.

\bigskip
\bigskip
\bigskip
{\large \sf Methods}
\bigskip

{\sf Protein model}
\bigskip

We represent interactions in a protein in a $C_{\beta}$ approximation
--- each pair of amino acids interact if their $C_{\beta}$s
($C_{\alpha}$ in the case of Gly) are within the contact distance
7.5\r A \cite{Jernigan96}. The total potential energy of the protein
can be written as follows:
\be
E = \frac{1}{2} \sum_{i\neq j }^{N}
U(\sigma_i, \sigma_j) \Delta_{ij} \, ,
\label{eq:en}
\ee
where $N$ is the length of the protein, and $\sigma_i$ is the amino
acid type at the position $i = 1,\dots,N$. $U(\sigma_i, \sigma_j)$ is
the corresponding element of the matrix of pairwise interactions
between amino acids $\sigma_i$ and $\sigma_j$.  $\Delta_{ij}$ is the
element of the contact matrix that is defined to be 1 if there is a
contact between amino acids $i$ and $j$, and is 0 otherwise.

\bigskip
{\sf Six-letter code potential}
\bigskip

Due to the similarities in the properties of the 20 types of amino
acids one can classify these amino acids into 6 distinct groups:
aliphatic $\{AVLIMC\}$, aromatic $\{FWYH\}$, polar $\{STNQ\}$,
positive $\{KR\}$, negative $\{DE\}$, and special (reflecting their
special conformational properties) $\{GP\}$. We construct the
effective potential of interaction, $U_6(\hat{\sigma}_i,
\hat{\sigma}_j)$, between six groups of amino acids, $\hat{\sigma}$,
by computing the average interaction between these groups, i.~e.
\be
U_6(\hat{\sigma}_i, \hat{\sigma}_j) = \frac{1}{N_{\hat{\sigma}_i} 
N_{\hat{\sigma}_j}} \sum_{\sigma_k \in \hat{\sigma}_i\, 
,\, \sigma_l \in \hat{\sigma}_j} U_{20}(\sigma_k, \sigma_l)\, ,
\label{eq:u6}
\ee
where $\sigma$ denotes amino acids in the 20-letter representation,
and $U_{20}(\sigma_k, \sigma_l)$ is the 20-letter Miyazawa-Jernigan
matrix of interaction \cite{Miyazawa96}; $\hat{\sigma}$ denotes amino
acids in the 6-letter representation. $N_{\hat{\sigma}}$ is the number
of actual amino acids of type $\hat{\sigma}$, e.~g. for the aliphatic
group, $N_{\hat{\sigma}} = 6$.


\bigskip
\noindent
{\small We thank R. S. Dokholyan for careful reading of the manuscript
and H. Angerman, S. V. Buldyrev, E. Kussell, L. Li, J. Shimada for
helpful discussions.  NVD is supported by NIH postdoctoral
fellowship. EIS is supported by NIH.}


\begin{thebibliography}{10}

\bibitem{Branden98} Branden, C. \& Tooze, J., {\em Introduction to
protein structure}, Garland Publishing Inc., New York, 1998

\bibitem{Mirny99} Mirny, L.~A. \& Shakhnovich, E.~I., Universally
conserved positions in protein folds: reading evolutionary signals
about stability, folding kinetics and function. {\em J. Mol. Biol.}
{\bf 291}, 177--196 (1999)

\bibitem{Ptitsyn98} Ptitsyn, O.~B., Protein folding and protein
evolution: Common folding nucleus in different subfamilies of c-type
cytochromes? {\em J. Mol. Biol.} {\bf 278}, 655--666 (1998)

\bibitem{Mirny98} Mirny, L.~A., Abkevich, V.~I. \& Shakhnovich, E.~I.,
How evolution makes proteins fold quickly. {\em
Proc. Natl. Acad. Sci. U. S. A.} {\bf 95}, 4976--4981 (1998)

\bibitem{vanNuland98a} van Nuland, N.~A.~J., {\em et al.}, Slow
folding of muscle acylphosphatase in the absence of intermediates.
{\em J. Mol. Biol.}  {\bf 283}, 883--891 (1998)

\bibitem{vanNuland98b} van Nuland, N.~A.~J., {\em et al.}, Slow
cooperative folding of a small globular protein hpr. {\em
Biochemistry} {\bf 37}, 622--637 (1998)

\bibitem{Dokholyan00a} Dokholyan, N.~V., Buldyrev, S.~V., Stanley,
H.~E. \& Shakhnovich, E.~I., Identifying the protein folding nucleus
using molecular dynamics. {\em J.  Mol. Biol.} {\bf 296}, 1183--1188
(2000)

\bibitem{Sali94} Sali, A., Shakhnovich, E.~I. \& Karplus, M., Kinetics
of protein folding. A lattice model study for the requirements for
folding to the native state. {\em J. Mol. Biol.} {\bf 235}, 1614--1636
(1994)

\bibitem{Shakhnovich93} Shakhnovich, E.~I. \& Gutin, A.~M.,
Engineering of stable and fast folding sequences of model proteins.
{\em Proc. Natl. Acad. Sci. U. S. A.} {\bf 90}, 7195--7199 (1993)


\bibitem{Ramanathan94} Ramanathan, S. \& Shakhnovich, E.~I.,
Statistical mechanics of proteins with evolutionary ``selected''
sequences. {\em Phys. Rev. E} {\bf 50}, 1303--1312 (1994)


\bibitem{Finkelstein95} Finkelstein, A.~V., Gutin, A. \& Badretdinov,
A., Why are the same protein folds used to perform different
functions? {\em Proteins} {\bf 23}, 142--149 (1995)

\bibitem{Pande95} Pande, V.~S., Grosberg, A.~Yu. \& Tanaka, T.,
Freezing Transition of random heteropolymers consisting of arbitrary
sets of monomers. {\em Phys. Rev. E} {\bf 51}, 3381--3393 (1995)

\bibitem{Bowie91} Bowie, J.~U., Luthy, R. \& Eisenberg, D., A method
to identify protein sequences that fold into a known 3-dimensional
structure. {\em Science} {\bf 253}, 164--170 (1991)

\bibitem{Holm93} Holm, L. \& Sander, C., Protein structure comparison
by alignment of distance matrices. {\em J. Mol. Biol.} {\bf 233},
123--138 (1993)

\bibitem{Rost97} Rost, B., Protein structures sustain evolutionary
drift. {\em Folding \& Design} {\bf 2}, S19--S24 (1997)

\bibitem{Tiana00} Tiana, G., Broglia, R. \& Shakhnovich, E.~I., Hiking
in the energy landscape in sequence space: A bumpy road to good
folders. {\em Proteins: Struc. Func.  Genet.} {\bf 39}, 244--251
(2000)

\bibitem{Dodge98} Dodge, C., Schneider, R. \& Sander, C., The hssp
database of protein structure-sequence alignments and family profiles.
{\em Nucl. Acid Res.} {\bf 26}, 313--315 (1998)

\bibitem{Press89} Press, W.~H., Flannery, B.~P., Teukolsky, S.~A. \&
Vetterling, W.~T., {\em Numerical recipes}, Cambridge University
Press, Cambridge, 1989

\bibitem{Itzhaki95} Itzhaki, L.~S., Otzen, D.~E. \& Fersht, A.~R., The
structure of the transition-state for folding of chymotrypsin
inhibitor-2 analyzed by protein engineering methods --- evidence for a
nucleation-condensation mechanism for protein-folding. {\em
J. Mol. Biol.} {\bf 254}, 260--288 (1995)

\bibitem{Hamill00} Hamill, S.~J., Steward, A. \& Clarke, J., The
folding of an immunoglobulin-like greek key protein is defined by a
common-core nucleus and regions constrained by topology. {\em
J. Mol. Biol.} {\bf 297}, 165--178 (2000)

\bibitem{Lopez96} L\'opez-Hern\'andez, E. \& Serrano, L., Structure of
the transition state for folding of the 129 aa protein ChyY resembles
that of a smaller protein, CI-2. {\em Folding \& Design} {\bf 1},
43--55 (1996)

\bibitem{Jernigan96} Jernigan, R.~L. \& Bahar, I., Structure-derived
potentials and protein simulations. {\em Curr. Opinion Struc. Biol.}
{\bf 6}, 195--209 (1996)

\bibitem{Miyazawa96} Miyazawa, S. \& Jernigan, R.~L., Residue-residue
potentials with a favorable contact pair term and an unfavorable high
packing density term, for simulation and threading. {\em
J. Mol. Biol.} {\bf 256}, 623--644 (1996)

\bibitem{Bernstein77} Bernstein, F.~C., {\it et al.}, The Protein Data
Bank: a computer-based archival file for macromolecular structures.
{\em J. Mol. Biol.} {\bf 112}, 535--542 (1977)

\bibitem{Abola87} Abola, E.~E., Bernstein, F.~C., Bryant, S.~H.,
Koetzle, T.~F. \& Weng, J.  \newblock Protein Data Bank.  \newblock in
Allen, F.~H., Bergerhoff, G. \& Sievers, R., eds., {\em
Crystallographic Databases-Information Content, Software Systems,
Scientific Applications}, pp. 107--132. Data Commission of the
International Union of Crystallography, Cambridge 1987.

\end{thebibliography}

\begin{table}[htb]
\begin{tabular}{l|c|cc|cc}
Fold&$N_s$&\multicolumn{2}{c}{Representative protein} &
     Correlation coefficient &  \\
    &    &PDB code \protect\cite{Bernstein77,Abola87}&$N$ 
                   &$S_{MF}(k)$ vs. $S_{acr}(k)$& $T_{sel}$ \\\hline
Ig  & 51 & $1TEN$ & 89 & 0.63 &  0.34  \\
OB  & 18 & $1MJC$ & 69 & 0.69 &  0.19  \\
R   & 166& $3CHY$ & 128& 0.71 &  0.25  \\
\end{tabular}
\vspace*{0.5cm}
\caption{The values of the correlation coefficient $r$ for the linear
regression of $S_{MF}(k)$ versus $S_{acr}$ for Ig, OB, and R, folds
and the corresponding optimal values of the temperature $T=T_{sel}$.}
\label{t:1}
\end{table}

\end{document}